\documentclass[preprint,3p,12pt]{elsarticle}
\usepackage{bm}
\usepackage{ulem}
\usepackage{mathtools}
\usepackage{psfrag}
\usepackage{graphicx}
\usepackage{paralist}
\usepackage{color}
\usepackage{amsmath}
\usepackage{amssymb}
\usepackage{varioref}
\usepackage{fancyhdr}
\usepackage{amsfonts}
\usepackage{bm}
\usepackage{float}
\usepackage{booktabs}
\usepackage{multirow}
\usepackage{array}
\usepackage{epstopdf}
\usepackage{threeparttable}
\usepackage[numbers]{natbib}
\usepackage{enumitem}
\usepackage{tabularx}
\usepackage{caption,subcaption}
\usepackage{tcolorbox}
\usepackage{textcomp}
\usepackage{url}
\usepackage{tikz}
\usetikzlibrary{shapes,arrows,positioning}

\newcommand{\real}{\hbox{\rm I\kern-0.2emR}}
\renewcommand{\mathbf}{\bm}

\usepackage{natbib}

\makeatletter
\def\blfootnote{\xdef\@thefnmark{}\@footnotetext}
\makeatother

\journal{Journal of Ocean Engineering and Science}

\begin{document}

\title{{\Large \bf How Do Ice Shelves Calve?\\Peridynamic Modeling of Ice Shelf Fracture Driven by Wave Erosion,
Basal Melting,
and Buoyancy Flexure}}

\author{Ying Song ${}^{\dagger}$}

\author{Xuan Hu ${}^{\ddagger}$}

\author{Jingrui Xu ${}^{\mathsection}$}

\author{Keming Zhu ${}^{\ddagger}$}

\author{Yuan Zhang ${}^{\mathsection, \mathparagraph}$}

\author{Wenjun Lu ${}^{\mathsection \star}$}

\author{Shaofan Li ${}^{\ddagger, \star}$}

\address{
${}^{\dagger}$College of Ocean Science and Engineering, Shanghai Maritime University, Shanghai, 201306, China.\\
}
\address{
${}^{\ddagger}$Department of Civil and Environmental Engineering, University of California, Berkeley, 94720, USA.\\
}

\address{
${}^{\mathsection}$Norwegian University of Science and Technology, 7491 Trondheim, Norway.\\
}

\address{
${}^{\mathparagraph}$Naval Engineering University, Wuhan, 430033, China\\
}

\begin{abstract}

An ice shelf is a floating extension of a land-based ice sheet into the ocean. It plays a crucial role in slowing down the flow of land ice into the sea, thus stabilizing the ice sheet. However, this stabilizing effect can be weakened by ice calving, a process in which large fragments of ice detach from the ice shelf. Although ice calving is widely acknowledged as a major contributor to ice mass loss, and its frequency and magnitude are highly sensitive to the environmental forcing, the underlying physics-based mechanisms remain poorly understood, particularly under ocean wave actions.

\textcolor{black}{In this context, we developed a nonlocal peridynamics (PD) framework to model the ice calving process subjected to wave-induced frontal corrosion. The proposed physics-based PD framework enables investigation of the coupled effects of self-weight bending, buoyancy-induced foot loosening, and ice calving process. To authors' best knowledge, this work represents the first attempt to employ a physics-based peridynamics framework for simulating ice calving processes. Compared with conventional finite element methods (FEM), the PD framework naturally captures crack initiation, interaction, and propagation without the need for special numerical treatments, thereby providing a robust tool for simulating fracture phenomena under large deformations and long-term environmental loading.}

To quantitatively resolve fracture processes, we implement\textcolor{black}{ed} a static first Piola Kirchhoff virial stress formulation within the PD framework, allowing direct evaluation of stress concentration and energy release  at evolving crack tips. \textcolor{black}{Subsequently, the} model is rigorously validated through one-to-one comparisons with finite-element stress fields, analytical beam-theory solutions, and recent field observations of wave-driven ice-shelf failure reported
by Sartore et al. (2025). \textcolor{black}{After detailed validation work, an idealized hundred-meter-scale ice shelf model was established. The influence of wave erosion rate, shelf geometry, crack length, and crack-propagation velocity was systematically investigated.} The simulations reveal distinct fracture regimes and clarify how wave forcing modulates stress redistribution and fracture acceleration, providing new mechanistic insight into wave-induced calving dynamics.

Overall, this study establishes peridynamics as a powerful next-generation tool for ice-shelf fracture modeling, bridging the gap between laboratory-scale fracture mechanics and large-scale cryospheric processes. By explicitly resolving calving initiation and evolution under ocean-wave forcing, the proposed framework substantially enhances predictive capability for ice-shelf stability and offers a new pathway for improving calving parameterizations in ice-sheet and sea-level-rise models.
\end{abstract}


\blfootnote{$^{\star}$ Corresponding authors: wenjun.lu@ntnu.no, shaofan@berkeley.edu}
\maketitle

\section{Introduction}
\label{intro}

Ice calving is a key process controlling the mass balance of ice shelves and marine-terminating glaciers in polar regions. Ongoing climate warming enhances calving through a combination of oceanic and atmospheric 
forcing, including increased ocean heat flux, surface melting, altered wave conditions, and changes in ice buoyancy. 
These processes weaken ice shelves structurally, promote fracture, and increase the likelihood of large calving events.
As calving intensifies, ice shelves lose their ability to buttress inland ice, accelerating glacier flow
and contributing to long-term sea-level rise.

\begin{figure}[ht]
\begin{center}            				
\includegraphics[height=3.0in]{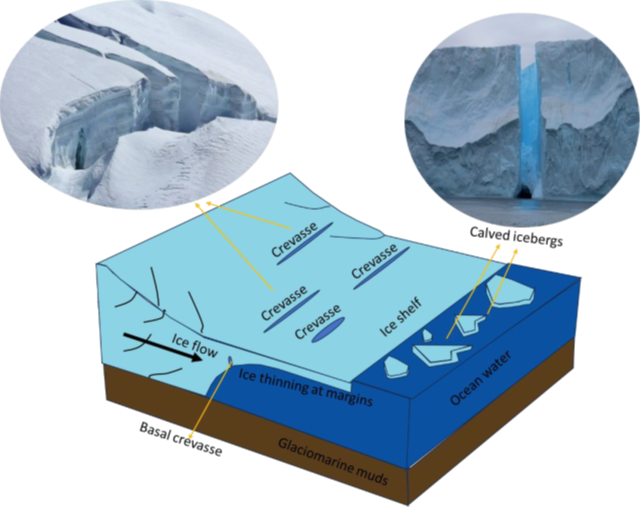}
\end{center}			
\caption{Sketch of evolution and calving of glacier and iceberg. Reproduced from 2D picture from
AntarcticGlaciers.org (https://www.antarcticglaciers.org/glacier-processes/glacier-types/ice-shelvessea-
ice-icebergs/)}
\label{fig:Background}
\end{figure}

Ice shelves are currently undergoing widespread thinning driven by atmospheric warming, enhanced basal melting, 
and evolving ocean circulation patterns 
\cite{amaral2020evaluation}. 
Thinning facilitates the initiation and growth of both surface and basal crevasses, 
which can widen under longitudinal tensile stresses and ultimately lead to calving, as illustrated in Fig.~\ref{fig:Background}.
Observational records indicate that calving events have become more frequent and larger 
in magnitude over recent decades, coinciding with accelerated ice mass loss and increasing concerns over future sea-level rise.

By modulating buttressing forces, calving plays a central role in ice-sheet dynamics. 
Recent estimates suggest that approximately $60 \% \pm 10 \% $of Antarctic ice shelves provide 
significant buttressing and are therefore susceptible to destabilization through fracture-driven processes 
\cite{Dow2018,lai2020vulnerability,Wang2025}. 
The loss of this buttressing can initiate rapid dynamic thinning and grounding-line retreat,
 amplifying the Antarctic 
ice sheets contribution to sea-level rise. As a result, uncertainties in calving physics
 remain a major limitation
in projections of future sea-level change, since calving governs not only the magnitude 
but also the rate of ice-sheet mass loss.

Ice-calving is a multi-physics phenomenon. It progresses and interacts with multiple physical processes:
sea-wave erosion at the ice-self front, basal melting, buoyancy flexure such as footloose, and
pre-existing cracks and stress concentrations. 
Satellite imagery, field measurements,
and laboratory studies show that calving is influenced by sea-wave erosion at the ice-shelf front,
basal melting, buoyancy-driven flexure such as footloose behavior, and the presence of pre-existing
fractures and stress concentrations \cite{Owczarek2024}.
A range of physical mechanisms have been identified to drive
ice shelf calving \cite{stern2019modeling}, including surface meltwater-driven hydrofracture \cite{slater2021calving},
basal melting induced by warm ocean currents \cite{buck2021flexural} as emerged as a potentially important but
relatively underexplored mechanism \cite{huth2023simulating}. When an ice shelf front is exposed to open water,
 ocean waves erode the submerged ice face, forming a basal notch that progressively deepens and widens.
 This process produces an overhanging ice geometry that is mechanically unstable and prone
  to collapse \cite{feshalami2025numerical}. Repeated cycles of wave erosion, notch growth,
  and collapse alter the stress state near the ice front and can trigger fracture and calving
  through the so-called foot-loosening mechanism.

Recent observational and theoretical studies have shown that wave-induced erosion at ice-shelf fronts 
can generate substantial bending moments and tensile stresses, particularly during energetic 
wave conditions \cite{sartore2025wave}. Satellite observations, field measurements, and idealized 
flexural models consistently indicate that these stresses concentrate near the ice front, 
preconditioning the ice for fracture and calving. However, most existing studies examine 
erosion kinematics or elastic flexural responses in isolation and do not explicitly resolve
 the initiation and propagation of fractures that ultimately govern calving behavior \cite{duddu2013nonlocal}.

Conventional continuum mechanics based numerical approaches, such as finite-element models, 
face inherent limitations in representing crack nucleation and growth, as they rely 
on spatial derivatives and fixed mesh connectivity \cite{krug2014combining}.  
An ice shelf is inherently a nonlocal porous medium, and the fracture physics of the ice shelf 
is essentially concern of crack propagation in a nonlocal porous medium \cite{Duddu2020,Schulson2022}. 
As a result, these models often require ad hoc fracture criteria or parameterizations that are not
 consistent with the failure theory of geological materials, limiting their ability to capture 
 the observed transition from wave-driven erosion and flexure to discrete fracture and calving.

To address these limitations, in this study, we adopt a nonlocal continuum mechanics model, peridynamics 
\cite{silling2000reformulation}, to study fracture and crevasse propagation.
The nonlocal fracture physics framework for ice \cite{song2023peridynamic}
is formulated using integral equations rather than localized spatial derivatives. 
Therefore,
Peridynamics naturally accommodates displacement discontinuities and provides a robust theoretical 
basis for simulating crack initiation, coalescence, and crevasse propagation without the need 
for ad-hoc fracture criteria or remeshing \cite{silling2000reformulation}. 
This makes it particularly well suited for modeling ice
shelf fracture under evolving geometries and loading conditions.

In this paper, we develop a special peridynamic model to investigate 
ice shelf fracture induced by wave-driven frontal erosion and bending. 
A systematic study for 
peridynamics constitutive modeling of sea ice has 
 been carried by the present authors \cite{song2020peridynamic,song2021peridynamic},
  which accounts for elastic-brittle behavior and strain-rate-dependent softening associated with creep processes. 
  The peridynamic ice model is applied to idealized ice shelf geometries subjected to progressive basal erosion 
  and wave-induced bending moments. Through numerical simulations, we analyze the evolution of stress fields, 
  crack initiation locations, crack propagation paths, and fracture characteristics under varying erosion rates and loading conditions.

The results provide new insight into the mechanical pathways linking continuous wave-induced erosion to discrete calving events and demonstrate the capability of peridynamics as a predictive tool for ice shelf fracture. This work contributes to improved understanding of wave-driven calving processes and offers a modeling framework that can be extended to more realistic ice shelf geometries and environmental forcings.

\section{Establishment of Ice Shelf Calving Model under Wave Erosion }
\label{sec:2}
\subsection{Problem Description}
\begin{figure}[ht]
  \begin{center}            				
  \includegraphics[height=3.0in]{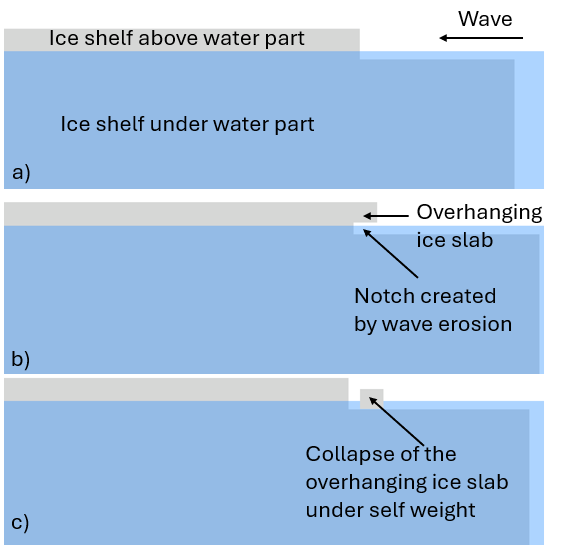}
  \end{center}			
  \caption{Idealised processes of wave erosion induced notch development and overhanging ice slab collapse}
  \label{fig:Icecollapse}
\end{figure}

This paper focuses on the front collapse events of ice shelves \textcolor{black}{, which is rarely explored in existing literature}. When an ice shelf is exposed to open water, wave-induced erosion leads to the formation of a notch at the water line (see Fig.~\ref{fig:Icecollapse}(b)). Over time, this notch progressively widens, resulting in the development of an overhanging ice slab. Continued exposure to wave action eventually causes the overhanging ice slab to collapse under its own weight, as illustrated in Fig.~\ref{fig:Icecollapse}(c). The collapsed ice slab is often about tens of meters in width and hundreds of meters in length. This process has been termed as front collapse in literature \cite{sartore2025wave}.

\begin{figure}[ht]
  \begin{center}            				
  \includegraphics[height=2.2in]{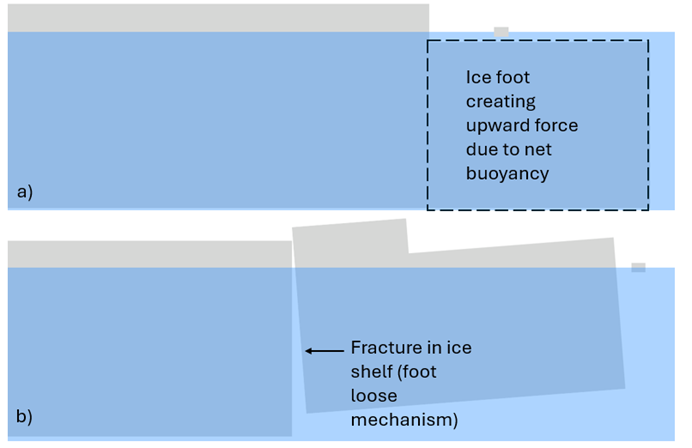}
\end{center}
  \caption{Foot loosing mechanism illustration}
  \label{fig:loosing}
\end{figure}

Following such a collapse, the remaining ice shelf hydrostatic equilibrium is further aggravated. The submerged front segment, referred to as the ``foot'' (see Fig.~\ref{fig:loosing}(a)), experiences an upward buoyant force due to wave interaction, inducing an upward flexural deformation at the ice shelf front. Repeated cycles of wave erosion and flexural stress accumulation lead to tensile stresses that, when exceeding the tensile strength of ice, initiate fracturing. This process is defined as the ``foot loosening'' mechanism of ice shelf disintegration. Compared to localized frontal collapse events, the foot-loosening mechanism operates at a substantially larger spatial scale, potentially extending over kilometers along the ice front and several hundreds of meters in the direction normal to the shoreline.

\begin{figure*}[ht]
  \begin{center}            				
  \includegraphics[height=2.0in]{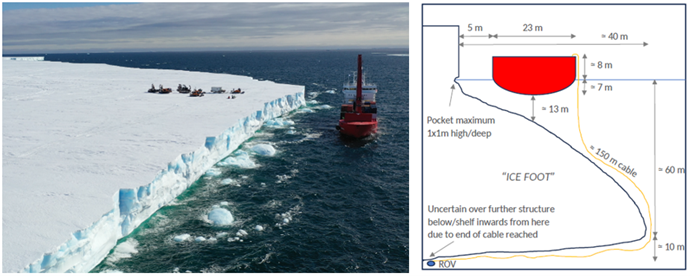}
\end{center}
  \caption{Left: A research vessel approaching the front of an Antarctic ice shelf. Right: Ice-shelf front profile measured using a remotely operated vehicle (ROV), showing a basal notch at the waterline and an underwater ice foot. The photographs and measurements shown are from different locations and are presented together for illustrative purposes only.}
  \label{fig:ROV}
\end{figure*}

\begin{figure*}[ht]
  \begin{center}            				
  \includegraphics[height=1.5in]{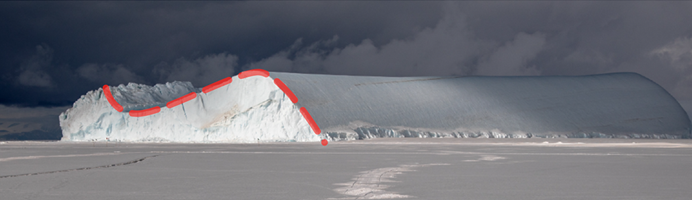}
\end{center}
  \caption{ Ice-shelf front geometry interpreted as the aftermath of a foot-loosening-type fracture, consistent with the conceptual mechanism shown in Figure 2(b) \cite{sartore2025wave}}
  \label{fig:front}
\end{figure*}

The mechanisms described above are supported by field observations. Fig.~\ref{fig:ROV} presents vessel-based measurements of the frontal geometry of an Antarctic ice shelf acquired by the Norwegian Polar Institute (NPI). Fig.~\ref{fig:front} shows an ice-shelf front interpreted as the aftermath of a foot-loosening-type failure. The red dashed line in Fig.~\ref{fig:front} conceptually corresponds to the ice-front profile shown in the right panel of Fig.~\ref{fig:ROV}.

These mechanisms involve coupled processes across multiple spatial and temporal scales, including wave-driven erosion, buoyancy-induced bending, and fracture evolution. Most existing modeling studies represent calving using strength- or stress-threshold criteria, in which fracture is not resolved explicitly. \textcolor{black}{For instance, the yield-stress-based calving formulation of Sartore et al.} \cite{sartore2025wave}. By contrast, a growing body of work incorporates fracture physics more directly through continuum damage mechanics\cite{duddu2013nonlocal,krug2014combining} or phase-field approaches approaches~\cite{sondershaus2023phase}, enabling explicit simulation of fracture processes within ice shelves. However, to the best of the authors' knowledge, dedicated fracture-resolving simulations of frontal collapse and foot-loosening mechanisms have not yet been reported.

This paper explores the use of peridynamics as a promising framework for simulating these relatively underexplored mechanisms, frontal collapse and foot-loosening with explicit resolution of fracture processes.

\begin{figure*}[ht]
  \begin{center}            				
  \includegraphics[height=2.2in]{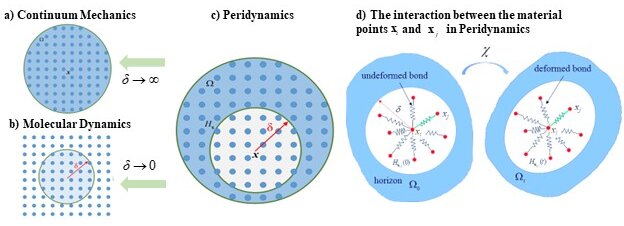}
\end{center}
  \caption{Illustration of the interaction mechanisms among material points in peridynamic nonlocal theory, continuum mechanics, and molecular dynamics}
  \label{fig:PD}
\end{figure*}


\subsection{\textcolor{black}{Introduction to} Peridynamic \textcolor{black}{Theory}}

\subsubsection{Rationale for Employing Peridynamics}

Conventional grid-based numerical methods, such as the finite element method, face inherent difficulties when simulating ice shelf fracture. These issues primarily arise from mesh distortion under large deformations and bending, as well as the failure of spatial derivatives at discontinuous crack interfaces. \textcolor{black}{In this context,} Peridynamics (PD) \cite{silling2000reformulation}, a novel meshless formulation, was introduced to effectively circumvent these limitations.

Peridynamics can be conceptually understood as a macroscale analogue to molecular dynamics, \textcolor{black}{see} Fig.~\ref{fig:PD}(a). In this framework, a continuous medium is discretized into a collection of material points. Each material point $\mathbf{x}$ is associated with a finite spherical region termed the horizon ${H}_{x}$ of radius $\delta$, as depicted in Fig.\ref{fig:PD}(c). Within this horizon, the point interacts with neighboring material points via nonlocal pairwise forces. This concept similar to interatomic bonds in molecular dynamics. The pairwise force density vector depends on the relative deformation between two connected points. As the continuum deforms, the length and orientation of these bonds change, thus altering the force density acting on each material point. The net force on a given material point is obtained by integrating all pairwise force densities over its horizon. This resultant force is then substituted into Newton's second law to determine the point's acceleration.

Benefiting from \textcolor{black}{its non-locality}, PD formulation naturally accommodates displacement discontinuities \textcolor{black}{without raising any singularity issues}, making it advantageous for multi-scale crack initiation and propagation problems. However, the nonlocal integral formulation complicates the direct definition and computation of stress, a concept central to classical mechanics. While a peridynamic stress tensor has been \textcolor{black}{proposed in its early development} \cite{silling2008convergence}, its complex triple integral form hinders practical application. \textcolor{black}{In 2022, Li et al.~\cite{li2022peridynamic} demonstrated the equivalence between peridynamic stress and static first Piola-Kirchoff virial stress, which eventually enables us to compute stress in a straightforward and efficient way.}

\subsubsection{Constitutive Modeling of Ice within the Bond-based Peridynamic Framework}

The mechanical behavior of ice under stress is critical for accurately simulating fracture. This section details the constitutive model developed for ice within the Bond-based peridynamic framework.

Peridynamics characterizes the constitutive behavior of materials through interaction forces between material points. The magnitude of the interaction force between material points ${{\mathbf{x}}_{i}}$ and ${{\mathbf{x}}_{j}}$ is modeled using a spring-like ``bond'' force, which is expressed as a function of position and displacement vectors:

\begin{equation}\label{eq.1}
\mathbf{f}({{\mathbf{\eta }}_{ij}},{{\mathbf{\xi}}_{ij}},t)=\mathbf{f}({{\mathbf{x}}_{i}},{{\mathbf{x}}_{j}})\mathbf{=}f({{\mathbf{x}}_{i}},{{\mathbf{x}}_{j}})\mathbf{n}~,
\end{equation}
\textcolor{black}{where $\mathbf{\xi}_{ij}:=\mathbf{x}_j - \mathbf{x}_i$ denotes the bond vector between $\mathbf{x}_i$ and $\mathbf{x}_j$, $\mathbf{\eta}_{ij}:=\mathbf{u}_j-\mathbf{u}_i$ represents the relative displacement between the two material points, $t$ is the time in PD formulation, and $\mathbf{n}$ means the corresponding unit vector of $\mathbf{\xi}$.}

The constitutive equation can be expressed by using the bond elongation ratio:
\begin{equation}\label{eq.2}
f({{\mathbf{x}}_{i}},{{\mathbf{x}}_{j}})\mathbf{=}
\begin{dcases}
    c\cdot\dfrac{\lvert\mathbf{\xi}_{ij}+\mathbf{\eta}_{ij}\rvert-\lvert\mathbf{\xi}_{ij}\rvert}{\lvert\mathbf{\xi}_{ij}\rvert}=cs, & s\le s_0\\
    0, & s > s_0
\end{dcases}~~,
\end{equation}
\textcolor{black}{where $c$ represents the micro-modulus of the bond, $\lvert\mathbf{\xi}\rvert=\xi, \lvert\mathbf{\eta}\rvert=\eta$ refer to the scalar magnitude of corresponding vectors $\mathbf{\xi}$ and $\mathbf{\eta}$, and $s:=\left(\lvert\mathbf{\xi}_{ij}+\mathbf{\eta}_{ij}\rvert-\lvert\mathbf{\xi}_{ij}\rvert\right)/\lvert\mathbf{\xi}_{ij}\rvert$ denotes the bond elongation ratio.} The micro-volume potential energy is defined as the interaction between two endpoints, denoted as $c({{\mathbf{x}}_{i}},{{\mathbf{x}}_{j}})$. The energy density per ``bond'' is referred to as the \sout{microscopic potential}\textcolor{black}{micro-potential} energy:

\begin{equation}\label{eq.3}
\omega =\frac{1}{2}c{{s}^{2}}\xi
\end{equation}

Consequently, the strain energy density function in the peridynamic formulation can be expressed as:

\begin{equation}\label{eq.4}
{{\text{W}}^{\text{PD}}}=\frac{1}{2}\int_{{{H}_{\mathbf{x}}}}{\frac{1}{2}cs^2\xi}d{{V}_{\xi }}=\frac{\pi cs^2\delta}{4}
\end{equation}

To further establishing the relationship between material point deformation and material constants, requiring that the peridynamic strain energy density function be consistent with conventional elasticity theory:

\begin{equation}\label{eq.5}
{{\text{W}}^{\text{CCM}}}=\frac{1}{2}\mathbf{\varepsilon }:\mathbb{C}:\mathbf{\varepsilon }
\end{equation}

\textcolor{black}{Combining Eq. (\ref{eq.4}), Eq. (\ref{eq.5}) and introducing $E$ to denote the material Young's modulus, the micro-modulus of the bond can be calibrated}:

\begin{equation}\label{eq.6}
c=\begin{dcases}
    \frac{12E}{\pi\delta^4}, & \mathrm{3D}\\
    \frac{9E}{\pi t \delta^3}, & \mathrm{plane~stress}\\
    \frac{48E}{5\pi t \delta^3}, & \mathrm{plane~strain}
\end{dcases}
\end{equation}

\begin{figure}[ht]
  \begin{center}            				
  \includegraphics[height=1.5in]{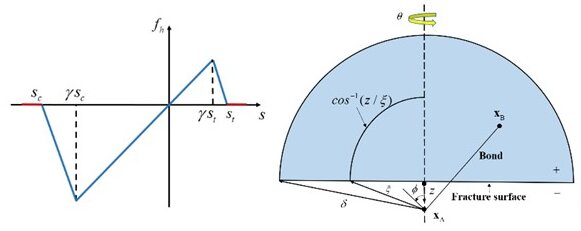}
\end{center}
  \caption{Illustration of the ice constitutive model and the fracture surface in bond-based peridynamics}
  \label{fig:Constitutive}
\end{figure}

In the standard PMB model, the bond force increases linearly with stretch $s$ until a critical stretch ${{s}_{0}}$ is reached, at which point the bond breaks instantaneously, simulating elastic-brittle behavior. However, the mechanical response of ice under tension involves a more complex transition prior to brittle fracture, it involves a fracture process zone (FPZ) characterized by distributed microcracking and damage accumulation ahead of the main crack tip. This leads to a gradual loss of load-carrying capacity before complete separation. To capture this essential quasi-brittle behavior, we introduce a strain-softening modification to the bond force-stretch law. As illustrated in Fig.\ref{fig:Constitutive}, within the framework of linear elastic fracture mechanics, ice fracture formation is predominantly caused by tensile failure, since the compressive strength of ice is approximately three to four times greater than its tensile strength.

A history-dependent scalar function is introduced as follows,
\begin{equation}\label{eq.def_mu_bond}
{{\mu }_{h}}({{\mathbf{\xi }}_{ij}},{{\mathbf{\eta }}_{ij}},{{\theta }_{ij}},t)\text{=}\left\{ \begin{matrix}
   \begin{matrix}
   1 & \begin{matrix}
   \begin{matrix}
   \begin{matrix}
   {} & {}  \\
\end{matrix} & {}  \\
\end{matrix} & {}  \\
\end{matrix}
\begin{matrix}
   , & s\in [0,\gamma {{s}_{0}}]  \\
\end{matrix}  \\
\end{matrix}  \\
\\
   \begin{matrix}
   \displaystyle \frac{{{s}_{0}}-(s-\alpha {{\theta }_{ij}})}{{{s}_{0}}-\gamma {{s}_{0}}} & \begin{matrix}
   , & s\in [\gamma {{s}_{0}},{{s}_{0}}]  \\
\end{matrix}  \\
\end{matrix}  \\
\\
   \begin{matrix}
   0 & \begin{matrix}
   \begin{matrix}
   \begin{matrix}
   {} & {}  \\
\end{matrix} & {} & {}  \\
\end{matrix}, & s\in [{{s}_{0}},+\infty ]  \\
\end{matrix}  \\
\end{matrix}  \\
\end{matrix} \right.
\end{equation}
The force density function between material points can be reformulated as follows:

\begin{equation}\label{eq.def_f_new}
f({{\mathbf{\xi }}_{ij}},{{\mathbf{\eta }}_{ij}},{{\theta }_{ij}},t)\text{=}cs\cdot{{\mu }_{h}}({{\mathbf{\xi }}_{ij}},{{\mathbf{\eta }}_{ij}},{{\theta }_{ij}},t)
\end{equation}

This formulation defines three distinct phases: In the linear elastic phase ($s<{\gamma{s}_{0}}$), representing the reversible elastic deformation of intact ice; After exceeding a yield stretch (${\gamma{s}_{0}}\leq{s}<{{s}_{0}}$), the bond force decreases linearly with increasing stretch. It captures the quasi-brittle fracture characteristics and the associated inelastic damage evolution within the FPZ under monotonic or cyclic loading. The model behavior is rate-independent in its current form; the parameters
$\gamma$ and ${{s}_{0}}$ can be calibrated to reflect different loading rates if needed; When the bond breaks irreversibly (${s}\geq{{s}_{0}}$), the bond is permanently broken, representing a microcrack. Macroscopic crack emergence and propagation are naturally represented by the collective failure of bonds.

When a bond fractures, the critical elongation ratio ${{s}_{0}}$ can be derived from the perspective of energy release rate. The material is first divided into two components along the damage surface, where all bonds connecting material points across this surface are fully fractured. The energy released per unit area of the damage surface, defined as the energy release rate ${{G}_{0}}$ required to separate the material into two parts, can be determined from material properties. This energy release rate ${{G}_{0}}$, which drives the fracture of bonds across each unit area of the damage surface, see Fig.~\ref{fig:Constitutive}, can be given by the following expression:

\begin{equation}\label{eq.9}
{{G}_{0}}=\int_{0}^{\delta }{\int_{0}^{2\pi }{\int_{z}^{\delta }{\int_{0}^{{{\cos }^{-1}}z/\xi }{(\frac{\xi }{2}c\left| \xi  \right|s_{0}^{2}){{\xi }^{2}}}}}}\sin \phi d\phi d\xi d\theta dz\\
\text{=}\frac{1}{2}cs_{0}^{2}\left( \frac{{{\delta }^{5}}\pi }{5} \right)
\end{equation}

By solving the integral equation, the energy per unit area of the damage surface can be determined. Given that this value is derived from material properties, and under the assumptions that the damage surface is fully separated and no additional energy dissipation occurs at the crack tip. The energy release rate ${{G}_{0}}$ quantifies a material's resistance to crack propagation. The fracture toughness coefficient ${{K}_{I}}$ is a key parameter characterizing ice strength. Empirical values of this coefficient can be determined from comprehensive experimental studies, and the critical elongation stretch ${{s}_{0}}$, expressed as follows:

\begin{equation}\label{eq.10}
{{\text{s}}_{0}}\text{=}\sqrt{\frac{5\pi {{G}_{0}}}{18E\delta }}=\sqrt{\frac{5\pi K_{I}^{2}}{18{{E}^{2}}\delta }}
\end{equation}

\subsubsection{Non-local First Piola-Kirchhoff Stress Calculation}

\textcolor{black}{In classical continuum mechanics, stress, specifically the divergence of stress, is a key component of equations of motion.} However, peridynamics adopts an integral form of the equations of motion to inherently handle displacement field discontinuities, which renders the classical stress concept inapplicable. To re-establish the constitutive relationship between stress and force within the peridynamic framework and achieve consistency with classical theories, it is necessary to develop a stress definition suitable for nonlocal theoretical systems.

Lehoucq and Silling \cite{silling2008convergence} first proposed the concept of the peridynamic stress tensor, proving that its divergence is equivalent to the net peridynamic force density acting on a material point, thus formally restoring consistency with classical equations of motion. However, this original definition involves complex triple integrals, making its computation extremely cumbersome and limiting its application in large-scale numerical simulations.

To address this challenge, Li et al.\cite{li2022peridynamic} demonstrated, based on Noll's nonlocal continuum mechanics, the Hardy-Murdoch continuumization process, and the Irving-Kirkwood statistical mechanics framework, that peridynamic stress is mathematically equivalent to the static first Piola-Kirchhoff Virial stress and derived a simplified expression suitable for numerical computation. This section will systematically elaborate on this stress calculation theory and establish its coupling framework with the ice constitutive model.

Consider a discrete system consisting of N particles, each with volume ${{V}_{i}}$ and position ${{\mathbf{X}}_{i}}$. According to the Virial theorem in statistical mechanics, the macroscopic stress of the system can be expressed as the sum of contributions from all internal interaction forces:

\begin{equation}\label{eq.11}
{{\Re }_{\text{Viral}}}=\frac{1}{\Omega}\sum\limits_{i<j}^{{}}{{{\mathbf{f}}_{ij}}}\otimes {{\mathbf{\xi}}_{ij}}
\end{equation}
where ${{\mathbf{f}}_{ij}}$ is the force exerted by particle j on particle i,  ${{\mathbf{\xi}}_{ij}}={{\mathbf{x}}_{j}}-{{\mathbf{x}}_{i}}$ is the relative position vector between the two particles, and $\Omega$ is the volume.

To generalize this discrete expression to a continuum, we introduce the Hardy-Murdoch continuumization 
process. Define a weighting function $\omega (\mathbf{r})$satisfying the normalization condition 
$\int_{H}{\omega (\mathbf{r})dV=1}$ and approaching
the delta function as $\left\| \mathbf{r} \right\|\to 0$. Through this weighting function, 
discrete particle forces can be ``smeared'' into
a continuous force density field as follows,

\begin{equation}\label{eq.12}
f(\mathbf{x},\mathbf{{x}'})=\sum\limits_{i=1}^{{{N}_{x}}}{\sum\limits_{j=1,j\ne i}^{{{N}_{x}}}{{{\mathbf{t}}_{ij}}\omega ({{\mathbf{x}}_{i}}-\mathbf{x})\delta (({{\mathbf{x}}_{J}}-{{\mathbf{x}}_{I}})}}-(\mathbf{{x}'}-\mathbf{x}))
\end{equation}
where ${{\mathbf{t}}_{ij}}$ is the force density vector acting on material point ${{\mathbf{x}}_{i}}$ from ${{\mathbf{x}}_{j}}$.

Based on Noll's (1955) nonlocal continuum mechanics theory, for a force density function satisfying the antisymmetric condition $\mathbf{f}(\mathbf{{x}'},\mathbf{x})=-\mathbf{f}(\mathbf{x},\mathbf{{x}'})$, we can define the nonlocal first Piola-Kirchhoff stress:

\begin{equation}\label{eq.13}
{{\Re }_{\text{Noll}}}(\mathbf{x}):=-\frac{1}{2}\int_{{{R}^{3}}}{\int_{0}^{1}{\mathbf{f}}}(\mathbf{x}+\alpha \mathbf{\xi},\mathbf{x}-(1-\alpha )\mathbf{\xi})\otimes \mathbf{\xi}d\alpha d{{V}_{\mathbf{\xi}}}
\end{equation}

This stress tensor satisfies the important property: its divergence equals the net peridynamic force density.

\begin{equation}\label{eq.14}
\nabla \cdot {{\Re }_{Noll}}(\mathbf{x})=\int_{{{H}_{x}}}{\mathbf{f}(\mathbf{{x}'},\mathbf{x})}d{{V}_{{{x}'}}}
\end{equation}

Substituting Eq.\ref{eq.11} into Eq.\ref{eq.12} and utilizing the sifting property of the delta function, we obtain:
\begin{equation}\label{eq.15}
\Re (\mathbf{x})=\frac{1}{2}\sum\limits_{i=1}^{{{N}_{x}}}{\sum\limits_{j\ne i}^{{N_x}}{{{\mathbf{t}}_{ij}}\otimes ({{\mathbf{x}}_{j}}-{{\mathbf{x}}_{i}})}}{{B}_{ij}}(\mathbf{x})
\end{equation}
where the bond function ${{B}_{ij}}(\mathbf{x})$ is introduced, defined as:
\begin{equation}\label{eq.16}
{{B}_{ij}}(\mathbf{x})=\int_{0}^{1}{\omega (\alpha ({{\mathbf{x}}_{j}}-{{\mathbf{x}}_{i}})+{{\mathbf{x}}_{i}}-\mathbf{x})}d\alpha
\end{equation}

\begin{figure}[ht]
  \begin{center}            				
  \includegraphics[height=2.0in]{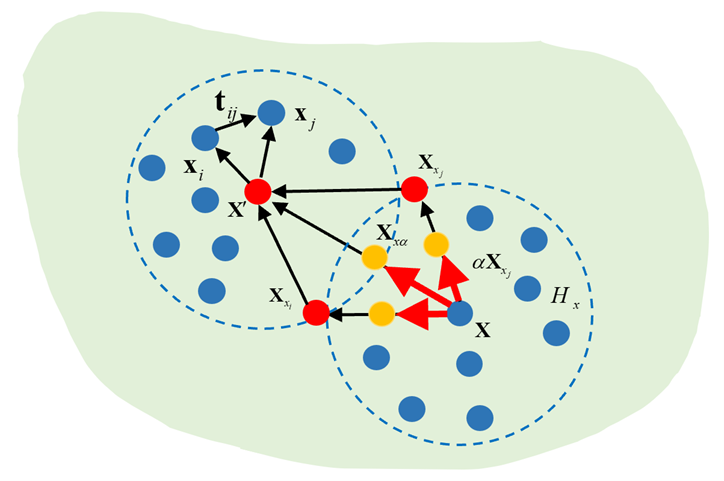}
\end{center}
  \caption{Illustration of the bond integration variable ${{\mathbf{x}}_{x\alpha }}$}
  \label{fig:bond}
\end{figure}

The bond function characterizes the weight of contribution of particle pair to the stress at spatial point $\mathbf{x}$, with its physical meaning illustrated in Fig.\ref{fig:bond}.

If ${{\mathbf{x}}_{i}},{{\mathbf{x}}_{k}}\in {{H}_{\mathbf x}}$, we can get the following equation as is illustrated in Fig.~\ref{fig:bond}.
\begin{equation}\label{eq.17}
{\mathbf{x}_{x\alpha }}:=\alpha ({{\mathbf{x}}_{j}}-\mathbf{x})+(1-\alpha )({{\mathbf{x}}_{i}}-\mathbf{x})\in {{H}_{x}}
\end{equation}

In practical computations, the radial step function is often employed as the weighting function:
\begin{equation}\label{eq.19}
\omega (\mathbf{\xi})=\left\{ \begin{matrix}
   \begin{matrix}
   \frac{1}{{{\Omega }_{x}}}, & \left\| \mathbf{\xi} \right\|<\delta   \\
\end{matrix}  \\
   \begin{matrix}
   0, & otherwise  \\
\end{matrix}  \\
\end{matrix} \right.
\end{equation}

With this choice, for all particle pairs located within the horizon of point $\mathbf{x}$, we have ${{B}_{ij}}(\mathbf{x})=\frac{1}{{{\Omega }_{x}}}$. Consequently, the stress formula simplifies to:
\begin{equation}\label{eq.20}
\Re (\mathbf{x})=\frac{1}{2{{\Omega }_{x}}}\sum\limits_{i=1}^{{{N}_{x}}}{\sum\limits_{j=1,j\ne i}^{{{N}_{x}}}{{{\mathbf{t}}_{ij}}}}\otimes ({{\mathbf{x}}_{j}}-{{\mathbf{x}}_{i}})
\end{equation}
Where, $\Re ({{\mathbf{x}}})$ is the first Piola-Kirchhoff stress tensor at point $\mathbf{x}$, ${{\Omega }_{x}}$ is the volume of the horizon centered at $\mathbf{x}$, $N_x$ is the number of particles within the horizon, ${{\mathbf{t}}_{ij}}$is the force density vector exerted on particle ${{x}_{i}}$ by particle ${{x}_{j}}$.

The peridynamic Virial stress formula has a clear physical meaning: the macroscopic stress at a point equals the average of the force-arm tensor products of all microscopic ``bonds'' within its horizon.

Substituting the ice constitutive model established in Section 2.2.2 into Eq.\ref{eq.13}, the force vector ${{\mathbf{t}}_{IJ}}$ can be specifically expressed as:
\begin{equation}\label{eq.21}
{{\mathbf{t}}_{ij}}=\mathbf{f}_{ij}(\mathbf{\eta }_{ij},\mathbf{\xi }_{ij}) \cdot {{V}_{i}}\cdot {{V}_{j}}
\end{equation}
Where ${{V}_{i}}$ and ${{V}_{j}}$ are the volumes associated with particles ${\mathbf{x}_{i}}$ and ${\mathbf{x}_{j}}$, respectively,
 and the pairwise force density in Eq. (\ref{eq.def_f_new}).

In general, the peridynamic first Piola-Kirchhoff virial stress tensor $\Re $ can be converted to the weight static Virial stress via:
\begin{equation}\label{eq.24}
\Re (\mathbf{x})=\frac{1}{2{{\Omega }_{x}}}\sum_{i=1}^{{{N}_{x}}}{\sum_{j=1,j\ne i}^{{{N}_{x}}}\left({{\mathbf{f}}_{ij}}\otimes ({{\mathbf{x}}_{j}}-{{\mathbf{x}}_{i}})\cdot V_i\cdot V_j\right)}
\end{equation}

\subsection{Framework and Assumption of Modeling Ice shelf fracture}
\label{sec:3}
Building upon the physical mechanisms described in Section 2.1, the fracture of ice shelves induced by wave erosion is here conceptualized within a unified PD modeling framework. The problem is characterized by the interaction between progressive geometric weakening at the ice front, buoyancy-induced bending, and fracture initiation and propagation within a brittle-creeping ice body. Rather than attempting to model the full complexity of a real ice shelf in a single step, the governing processes are decomposed into two idealized but physically representative scenarios.

The first scenario corresponds to frontal collapse, in which wave-driven erosion near the waterline produces an overhanging ice slab that ultimately fails under its own weight. In this case, fracture is primarily controlled by gravity-induced bending stresses and stress redistribution following the progressive loss of basal support. The second scenario represents the foot-loosening mechanism, where the buoyancy of the submerged ice foot (created by wave erosion) generates upward bending moments at the ice shelf front, leading to tensile stress accumulation, crack initiation, and large-scale fracture propagation. These two scenarios reflect distinct but sequential stages of wave-ice interaction observed in the field and discussed in Section 2.1.

Based on following described framework, the two scenarios are formulated as PD fracture problems of cantilever-like ice shelves subjected to gravity loading and bending moment, presented in Sections 2.3.1 and 2.3.2, respectively. The corresponding numerical implementations and results are presented in Sections 3.1 and 3.2.
\subsubsection{\textbf{Scenario 1 Front collapse}:  Fracture of an Elastic Cantilever Beam under Gravity}

\begin{figure}[ht]
  \begin{center}            				
  \includegraphics[height=1.5in]{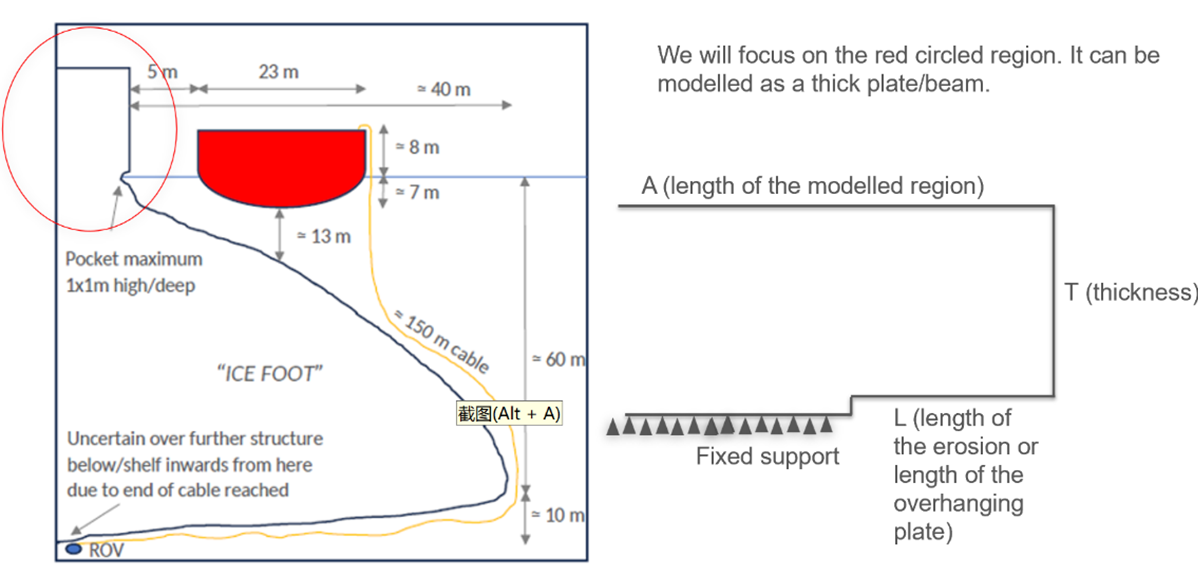}
\end{center}
  \caption{Description of the idealized ice shelf model under wave erosion}
  \label{fig:beam}
\end{figure}

To develop a mechanically tractable representation of the frontal collapse process outlined in Section 2.1, the ice shelf is idealized as a cantilever beam experiencing progressive loss of basal support due to wave-induced erosion, as depicted in Fig.\ref{fig:beam}. The beam possesses a length A, thickness T, and a unit width b. Its left end ($x = 0$) is treated as fixed under initial conditions, while the bottom is supported by uniformly distributed elastic foundations that represent buoyancy forces. Wave-driven erosion is simulated through the sequential removal of these basal supports. Key modeling assumptions include: (1) The ice is modeled as an isotropic, linearly elastic, and brittle material; (2) The removal of basal support caused by erosion is treated as quasi-static relative to elastic wave propagation in ice, reflecting the pronounced separation between the timescale of wave-driven notch growth (hours to days) and the timescale of elastic stress redistribution (seconds); and (3) The analysis is confined to plane strain conditions. Relevant material parameters for ice are provided in Table \ref{tab:load}, consistent with the existing data \cite{timco2010review,dempsey1991fracture}. Utilizing peridynamic (PD) theory, the mechanical response of the ice shelf subject to wave erosion is idealized, and a corresponding peridynamic model is formulated. The robustness of the proposed mechanical model is validated through comparisons with finite element simulations and available experimental test results.

\begin{table}
\centering
\caption{The parameters of ice}
\label{tab:load}
\begin{tabular}{lllll}
\hline
\multirow{2}{*}{Parameters} & \multirow{2}{*}{\begin{tabular}[c]{@{}l@{}}Unit\end{tabular}} & \multirow{2}{*}{\begin{tabular}[c]{@{}l@{}}Value\end{tabular}} \\
 &  &  &  &  \\ \hline
Elastic modulus $E$  & GPa & 9.31 &  \\ \hline
Density of  ice sheet $\rho_{i}$ & $kg/m^{3}$ & 897.6 &  \\ \hline
Poisson¡¯s ratio $\nu$ &  & 0.33 &  \\ \hline
Energy release rate ${{G}_{0}}$ & $J/m^{2}$ & 5 &  \\ \hline
Fracture toughness ${{K}_{I}}$  & $\text{kPa}\cdot {{\text{m}}^{1/2}}$ & 134 &  \\ \hline

\end{tabular}
\end{table}

In the application scenarios described above, the two-dimensional motion equations of peridynamics can be formulated to include terms representing body forces:
\begin{equation}\label{eq.25}
\rho ({{\mathbf{x}}_{i}})\mathbf{\ddot{u}}({{\mathbf{x}}_{i}},t)=\int\limits_{{{H}_{{{x}_{i}}}}}{\mathbf{f}({{\mathbf{\eta }}_{ij}},{{\mathbf{\xi }}_{ij}},t)d{{V}_{{\mathbf{\xi}_{j}}}}}+\mathbf{b}({{\mathbf{x}}_{i}},t)
\end{equation}

\textcolor{black}{The body force density $\mathbf{b}({{\mathbf{x}}_{i}},t)$ includes gravitational forces and support reactions.} The gravitational force is modeled as a body force uniformly distributed over all material points.
\begin{equation}\label{eq.26}
{{\mathbf{b}}_{gravity}}({{\mathbf{x}}_{i}},t)=\rho_{i} ({\mathbf{x}_{i}},t)\mathbf{g}
\end{equation}

This elastic foundation representation provides an effective mechanical abstraction of hydrostatic buoyancy, 
enabling controlled investigation of stress redistribution and fracture triggered by progressive buoyancy loss. 
The gradual removal of supports is implemented through a time-dependent reaction force function. The reaction force density is defined as:
\begin{equation}\label{eq.27}
{{\mathbf{b}}_{support}}({{\mathbf{x}}_{i}},t)=
\begin{dcases}
    (0,~{{k}_{s}}\cdot\alpha ({\mathbf{x}_{i}},t){{u}_{y}}({{\mathbf{x}}_{i}},t)), & \mathrm{if}~y=0~\mathrm{and~support~exists}\\
    (0,~0), & \mathrm{if~support~is~moved}
\end{dcases}
\end{equation}
Where ${{k}_{s}}$ is the support stiffness coefficient, and $\alpha ({\mathbf{x}_{i}},t)$ is the support existence function, defined as:
\begin{equation}\label{eq.28}
\alpha ({\mathbf{x}_{i}},t)=H(x-{{x}_{front}}(t))
\end{equation}
Here, $H(\bullet)$ denotes the Heaviside step function, and ${{x}_{front}}(t)$ represents the position of the erosion front, which propagates from right to left with time:
\begin{equation}\label{eq.29}
{{x}_{front}}(t)=L-{{v}_{e}}\cdot t
\end{equation}
Where ${{v}_{e}}$ is the erosion rate, and $x$ represents the position of the erosion front, which propagates from right to left with time.

This scenario primarily serves to validate the capability of the proposed PD framework to capture fracture initiation and propagation under gravity-dominated loading, providing a mechanical baseline for the more complex wave-induced bending scenario examined in Section 2.3.2.

\subsubsection{\textbf{Scenario 2 Foot loosing}: Fracture of Elastic Cantilever Beams under under net buoyancy}

To represent the foot-loosening mechanism described in Section 2.1, the ice shelf
is modeled as a floating elastic beam subjected to net buoyancy at the front,
with their bending behavior governed by the following governing equation:
\begin{equation}\label{eq.30}
B\frac{{{\partial }^{4}}\omega }{\partial {{x}^{4}}}+{{\rho }_{i}}h\frac{{{\partial }^{2}}\omega }{\partial {{t}^{2}}}=q(x,t)
\end{equation}
where $\omega$ denotes the deflection of the ice shelf, $B$ represents the bending stiffness,
and q(x, t) corresponds to the externally applied distributed load.

In the PD model, a bending moment $M$ is simulated by applying a force couple to nodes near the ice shelf front. The position $x = L$ is defined as the ice shelf front (free end), while $x = 0$ is set as the fixed end (corresponding to the grounding line). A bending moment M is applied at $x = L$ (the glacier front). The support represents the buoyant force from seawater, and removing the support simulates the loss of buoyancy due to seawater erosion. Buoyancy is applied as a distributed force on the bottom nodes. The variation of buoyancy with deflection is neglected, and hydrostatic buoyancy is applied as a distributed load on the bottom nodes. A net buoyant force is applied to each node at the base.
\begin{figure}[ht]
  \begin{center}            				
  \includegraphics[height=2.0in]{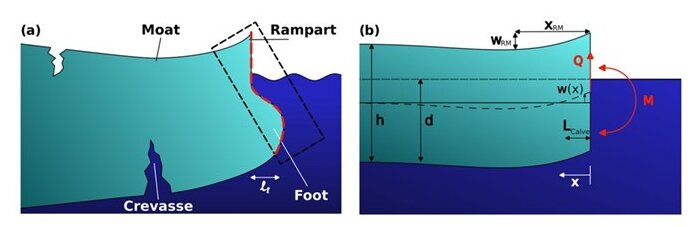}
\end{center}
  \caption{Foot-induced deformation of an ice shelf front (From reference~\cite{sartore2025wave})}
  \label{fig:shelf}
\end{figure}

Periodic bending moments induced by wave action at the ice shelf front are expressed as:
\begin{equation}\label{eq.31}
M(t)={{M}_{0}}\sin (2\pi ft)
\end{equation}
Herein, ${{M}_{0}}$ denotes the amplitude of the bending moment, while $f$ represents the wave frequency. This moment-based representation provides a mechanically equivalent description of wave-induced frontal bending, capturing the dominant effect of wave-ice interaction on stress accumulation and fracture initiation without explicitly resolving fluid-structure interaction. Although the applied bending moment is periodic, no explicit fatigue law is introduced; fracture emerges naturally from stress accumulation and bond failure within the peridynamic framework.

The base of the ice shelf is subject to a net buoyant force, and the net buoyancy per unit length is expressed as:
\begin{equation}\label{eq.32}
{{q}_{b}}=({{\rho }_{w}}-{{\rho }_{i}})gh
\end{equation}
where $\rho$ denotes the density of seawater.

In this scenario, buoyancy provides the background floating constraint, while fracture initiation and propagation are primarily driven by cyclic bending stresses induced by wave action at the ice shelf front.

\section{Numerical Results and Discussions}

This section presents numerical results obtained using the proposed PD framework.
Two representative scenarios are examined, corresponding to the frontal collapse and foot-loosening
mechanisms introduced in Section 2.3. Section 3.1 focuses on gravity-dominated fracture
as a baseline validation case, while Section 3.2 investigates wave-induced bending
fracture relevant to ice shelf front destabilization.
\subsection{Scenario 1: Front collapse}

The numerical configuration follows the modeling framework described in Section 2.3.1.
The ice shelf is idealized as a two-dimensional cantilever structure with the grounding line
fixed and buoyant support applied along the bottom surface. Wave-induced erosion is represented
by the gradual removal of basal supports from the ice front toward the interior, mimicking
the development of an overhanging geometry. The ice body is discretized into uniformly distributed
material points, and gravitational loading is applied as a body force. The erosion process
is assumed to be quasi-static relative to elastic wave propagation, allowing stress redistribution to occur prior to fracture.

\subsubsection{Numerical setup and objective}

To validate the proposed peridynamic model for ice shelf fracture, a representative
numerical experiment was conducted to simulate ice shelf cracking. Model parameters were
established based on typical physical properties of Antarctic ice shelves, as summarized in Table.\ref{tab:load}.
\begin{figure}[ht]
  \begin{center}            				
  \includegraphics[height=0.9in]{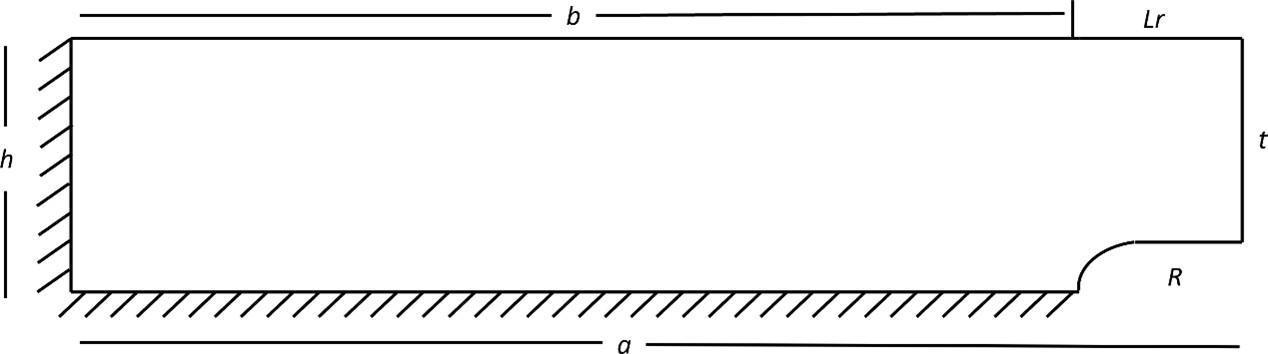}
  \end{center}		
  \caption{Geometric model for numerical simulation}
  \label{fig:geometic}
\end{figure}

The \sout{finite element}\textcolor{black}{geometric} model developed in this study is depicted in Fig. \ref{fig:geometic}, representing
a typical ice shelf overhang under environmental forcing. Key geometric parameters include the overhang
thickness $t$, the overhang length defined as ${{L}_{r}}=a-b$, and the transition fillet radius $R$,
which satisfies the relation $h-t = R$. This configuration is designed to simulate the realistic
morphological evolution of an ice shelf, wherein continuous wave erosion at the waterline gradually
carves a basal fillet. In order to characterize the progressive nature of this erosion,
the fillet radius $R$ is treated as a variable, with the dimensionless ratio $R/t$ varying
from 0.1 to 0.9. Fixed geometric parameters are set as follows: $a=100$ m, $b=90$ m, ${{L}_{r}}=10$ m, and $t=10$ m.

Regarding boundary conditions, the left vertical face and the bottom surface
are fully constrained to represent rigid attachment to the inland ice sheet.
A uniform gravitational acceleration is applied throughout the domain to account
for the self-weight of the ice shelf. All other boundaries remain free,
ensuring that the resulting stress and displacement fields arise solely
from the cantilevered mass and the local geometry of the fillet.

The primary objective of the numerical simulation is to evaluate the maximum
tensile stress on the upper surface and the total displacement magnitude as functions
of the $R/t$ ratio. To ensure solution reliability, a mesh convergence study
was conducted by comparing two mesh densities: a standard mesh with a global element
size of 0.25 m and local refinement of 0.05 m in the overhang region, and a finer mesh
with a global size of 0.125 m and local refinement of 0.025 m. The discrepancy
in peak tensile stress between the two meshes was approximately 3 percent,
while the variation in the total displacement magnitude was negligible, effectively approaching zero.
 These results confirm that the numerical solution has achieved a high degree of convergence. Consequently,
 the standard mesh configuration (global 0.25 m, local 0.05 m) was selected for all subsequent simulations,
 optimizing computational efficiency without compromising result accuracy.
\begin{figure*}[ht]
  \begin{center}            				
  \includegraphics[height=2.2in]{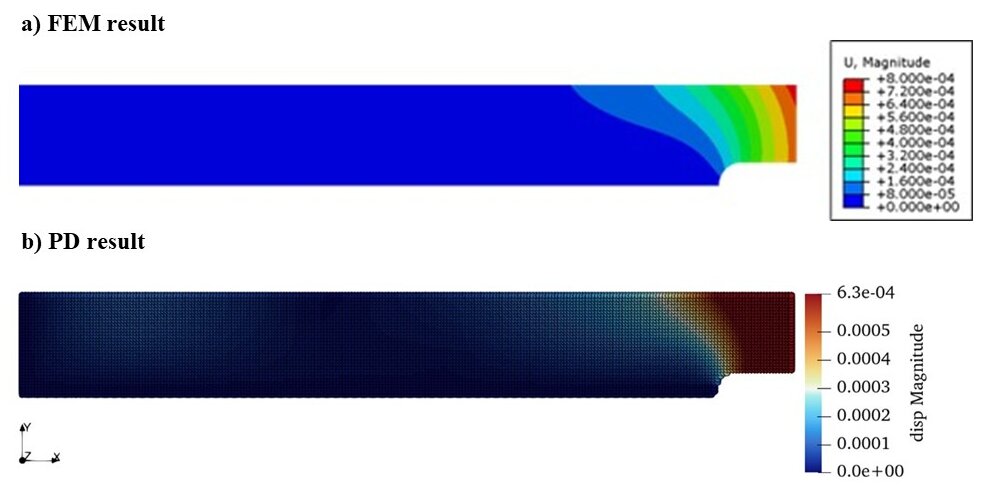}
  \end{center}		
  \caption{The comparison of the displacement result between FEM and PD}
  \label{fig:12}
\end{figure*}
\begin{figure*}[ht]
  \begin{center}            				
  \includegraphics[height=2.2in]{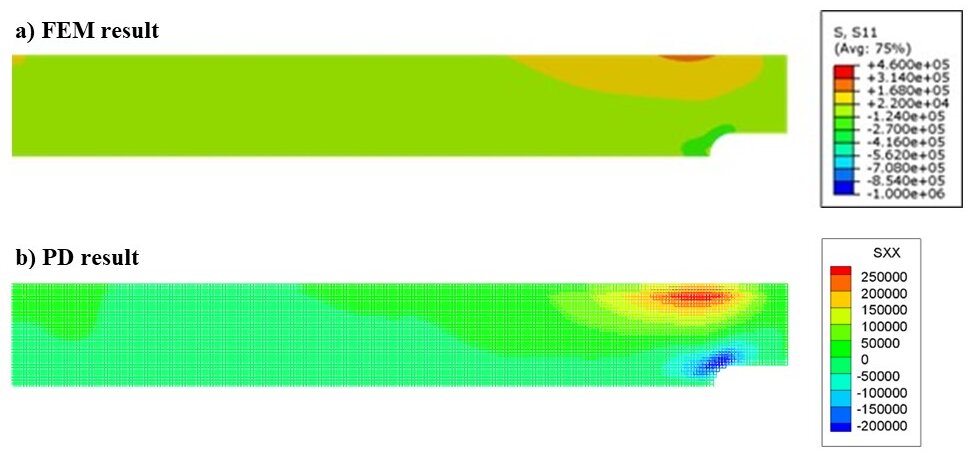}
  \end{center}		
  \caption{The comparison of the $S_{11}$ stress result between FEM and PD}
  \label{fig:13}
\end{figure*}


To rigorously validate the accuracy and reliability of the proposed peridynamic (PD) model in simulating the quasi-static mechanical response of an ice shelf prior to fracture, a direct quantitative comparison was conducted against results from a conventional Finite Element Method (FEM) simulation. Both models were constructed with identical geometric parameters (as defined in Fig.\ref{fig:geometic} and Section 3.1.1), material properties (Table \ref{tab:load}), boundary conditions (fixed left end, elastic foundation support at the base),
and gravitational loading. This ensures a consistent benchmark for evaluating the PD formulation's
 capability to replicate stress and displacement fields predicted by a well-established continuum mechanics approach.

\begin{figure}[h]
  \begin{center}            				
  \includegraphics[height=4.0in]{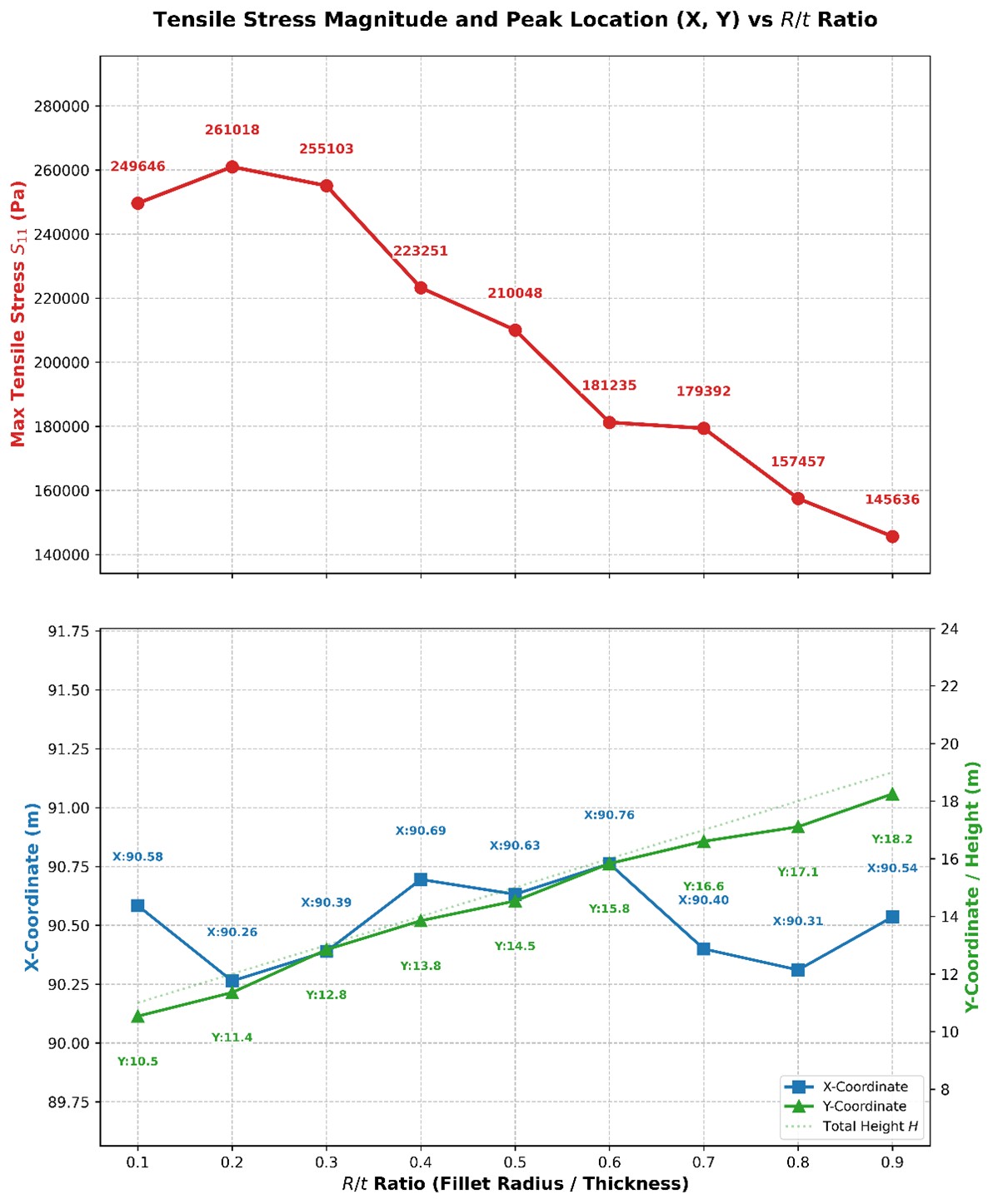}
  \end{center}		
  \caption{The quantitative relationship between the maximum tensile stress, its spatial coordinates, and the geometric ratio $R/t$}
  \label{fig:17}
\end{figure}

The comparative results for the representative case of $R/t = 0.3$ are presented in Fig.\ref{fig:12}-\ref{fig:13}. The corresponding distribution of the total displacement magnitude is shown separately in Fig.\ref{fig:17}. \textcolor{black}{A complete comparative data set encompassing all investigated $R/t$ ratios is provided in the Appendix.}

The displacement fields (see Fig.\ref{fig:12}) show excellent qualitative and quantitative agreement. Both models predict the characteristic cantilever bending pattern, with the maximum displacement localized at the free end of the overhang. This close match confirms that the PD discretization and force formulation correctly capture the global deformation mechanics under gravity-dominated loading.

The comparison of the \textcolor{black}{Cauchy stress component $\sigma_{11}$} (see Fig.~\ref{fig:13}) is particularly critical, as it governs tensile fracture initiation. Both models identify the same region of high stress concentration, the upper surface near the geometric transition between the grounded section and the overhang. The local derivative-based FEM can develop sharper stress gradients near geometric singularities, while the non-local integral formulation of PD naturally smooths such concentrations through its horizon-based force averaging. Importantly, both models concur on the location and magnitude of the critical tensile zone, which validates the PD model's ability to accurately identify potential fracture initiation sites.

This comprehensive comparison establishes that the PD framework faithfully replicates the elastic stress and displacement fields computed by a standard FEM approach for intact ice shelf geometries. The demonstrated agreement provides a solid foundation for confidence in the subsequent PD simulations of fracture propagation regime where the PD method offers a distinct advantage due to its inherent ability to model spontaneous crack nucleation and growth without recourse to special crack-tip elements or ad-hoc failure criteria, limitations inherent to the FEM for such problems.

Analysis of the displacement field reveals that the maximum displacement magnitude consistently localizes at the free end of the ice shelf overhang. This pattern is characteristic of a cantilever bending response.

Furthermore, simulation results across the entire range of $R/t$ ratios indicate that the overall displacement distribution remains largely invariant with respect to the fillet radius. This finding suggests that the change in basal geometry induced by wave erosion has a limited influence on the global bending deformation of the structure, as is shown in Fig.\ref{fig:17}. In contrast to the displacement field, the longitudinal tensile stress, demonstrates a pronounced sensitivity to the local transition geometry. As the data in Fig.\ref{fig:17} summarize, the peak tensile stress on the upper surface shows a slight initial fluctuation within the lower $R/t$ range (0.1 to 0.3), after which it undergoes a consistent monotonic decrease as the ratio increases to 0.9. The minor instability observed at very low $R/t$ values is likely attributable to numerical sampling errors inherent in meshing an extremely small fillet radius, rather than to a genuine mechanical response. For $R/t>0.3$, the trend is clear: progressive smoothing of the basal profile¡ªsimulated by an increasing fillet radius¡ªeffectively alleviates stress concentration at the geometric transition.

\begin{figure}[ht]
  \begin{center}            				
  \includegraphics[height=2.0in]{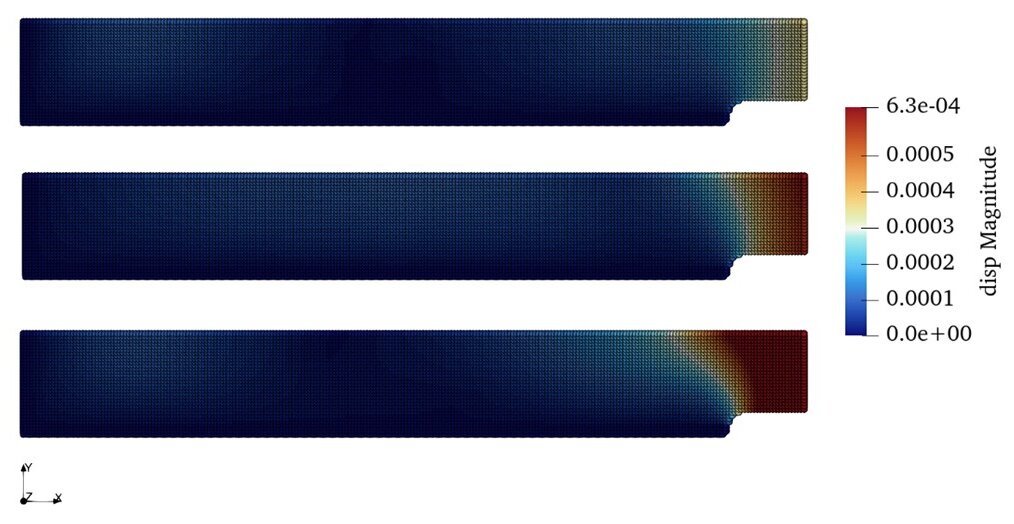}
  \end{center}		
  \caption{Peridynamics
  simulation results of the displacement evolution under gravity load before ice fracture occurred}
  \label{fig:18}
\end{figure}

\begin{figure}[ht]
  \begin{center}            				
  \includegraphics[height=2.4in]{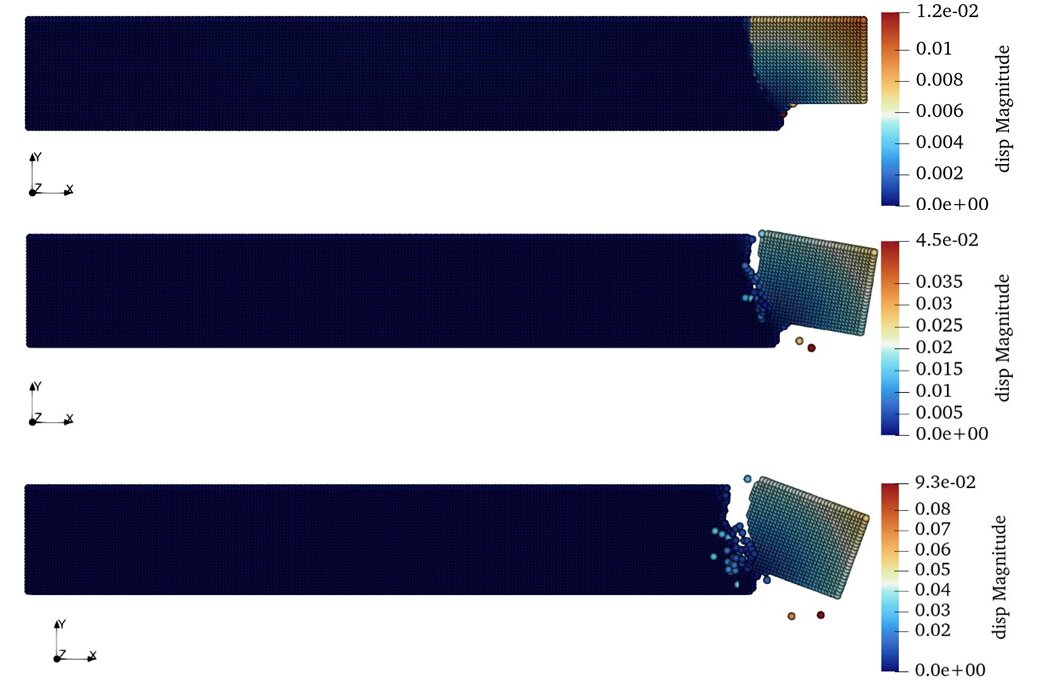}
  \end{center}		
  \caption{Peridynamics simulation results of the displacement evolution under gravity load after ice fracture occurred}
  \label{fig:19}
\end{figure}

\begin{figure}[ht]
  \begin{center}            				
  \includegraphics[height=2.4in]{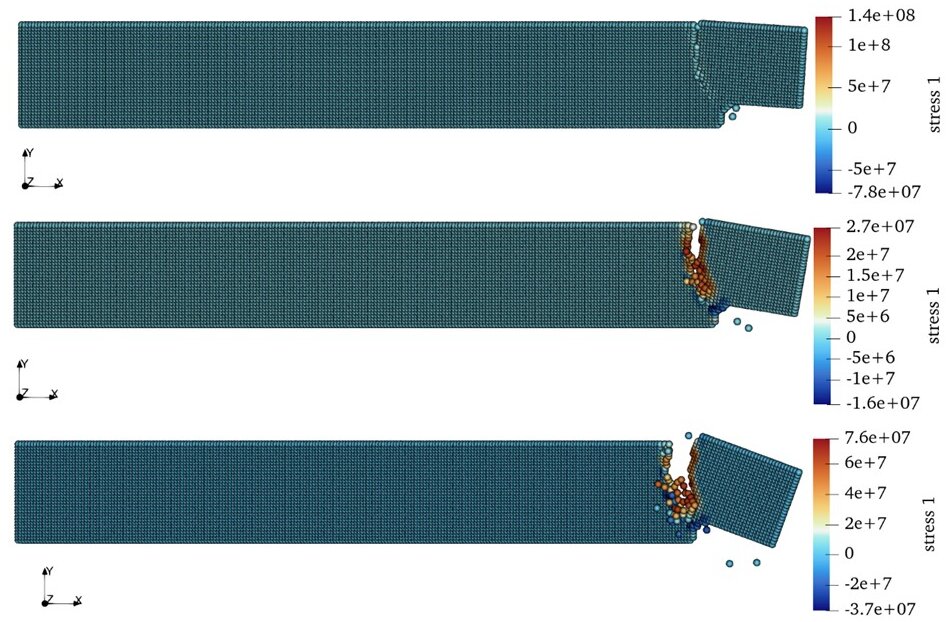}
  \end{center}		
  \caption{Peridynamics simulation results of the $\sigma_{11}$ stress evolution under gravity load after ice fracture occurred.}
  \label{fig:20}
\end{figure}

\begin{figure}[ht]
  \begin{center}            				
  \includegraphics[height=2.5in]{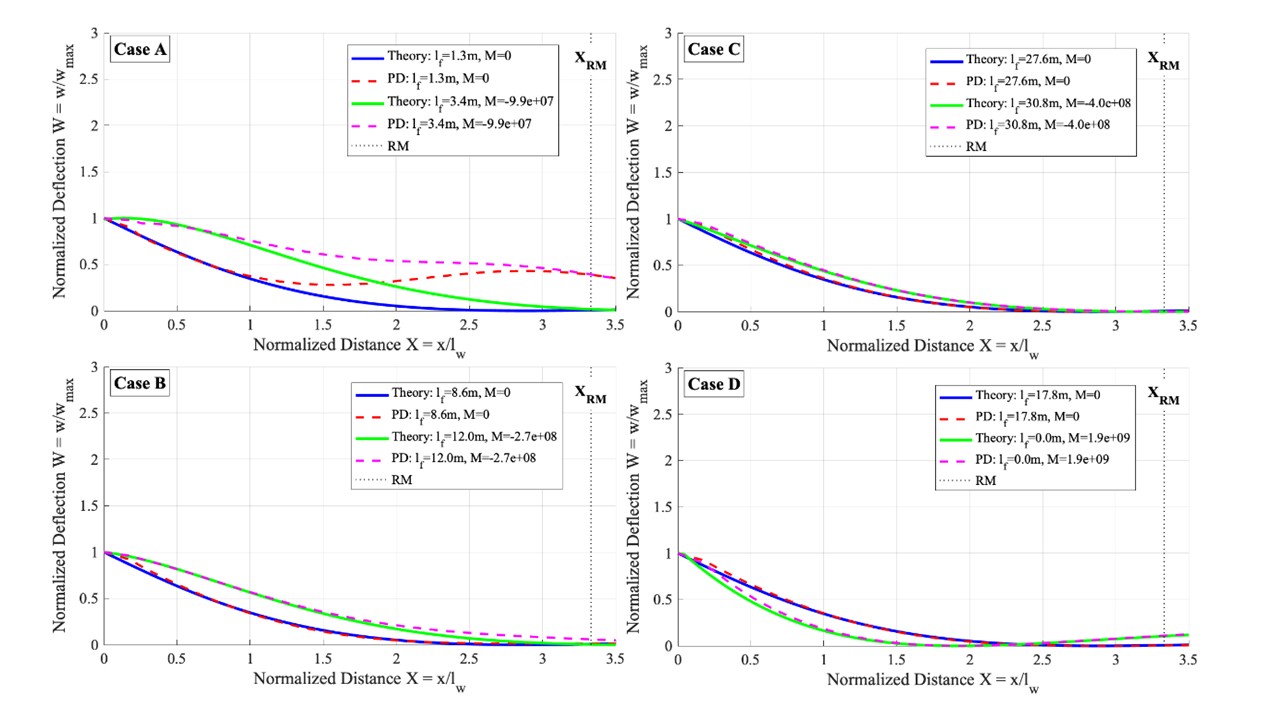}
  \end{center}	
  \caption{Validation of the PD model by comparing the numerical results with
   the classical beam theory \cite{sartore2025wave}}
  \label{fig:22}
\end{figure}


Once the tensile stress exceeds the strength of ice, fracture initiates within the overhanging slab.
Figs.\ref{fig:19}-\ref{fig:20} present the displacement and stress fields after fracture onset.
At this stage, discontinuities in displacement and stress become evident, and crack propagation leads
to rapid stress release in the fractured region. The crack path develops naturally within the PD framework
without predefining fracture locations or directions, demonstrating the ability of the model to explicitly
resolve fracture initiation and growth under gravity-dominated loading

\subsection{Scenario 2: Foot loosing}

This scenario investigates fracture initiation and propagation in an ice shelf
due to cyclic wave-induced bending moments, a process referred to as the ``foot loosening''
 mechanism, corresponding to the foot-loosening mechanism described in Section 2.1. In contrast to Scenario 1,
 where gravity dominates the mechanical response, fracture here is driven by cyclic bending moments applied at the ice shelf front,
  while buoyancy provides a background floating constraint.
 This configuration enables direct examination of how repeated wave-induced bending
  leads to stress accumulation, crack initiation, and large-scale fracture propagation.

Figure \ref{fig:22} presents a direct comparison between the peridynamic (PD) simulation results and the analytical elastic beam theory outlined in the reference study \sout{(Sartore et al., 2025)}\cite{sartore2025wave}. The PD results for Case A and Case C show excellent agreement with the theoretical predictions. The simulated ${x_{RM}}$ values (the horizontal distance from the ice front to the moat depression) remain stable across variations and align closely with the expected theoretical values, with associated uncertainties near zero. This high-fidelity match confirms that the PD framework correctly replicates the fundamental deformation pattern of a floating ice shelf subject to a buoyant foot, as described by the classical elastic beam solution. In contrast, Case B and Case D demonstrate a controlled, incremental increase in ${x_{RM}}$ with each successive parameter step. This pattern is not a discrepancy but a deliberate exploration of sensitivity. It accurately reflects how the theoretical model predicts changes in deformation scale when underlying parameters are systematically varied, such as flexural rigidity or the magnitude of the applied moment. The precise, low-error progression of the PD results through this parameter space demonstrates
that the numerical model correctly captures the scaling relationships inherent to the theoretical formulation.

\begin{figure*}[ht]
  \begin{center}            				
  \includegraphics[height=4.5in]{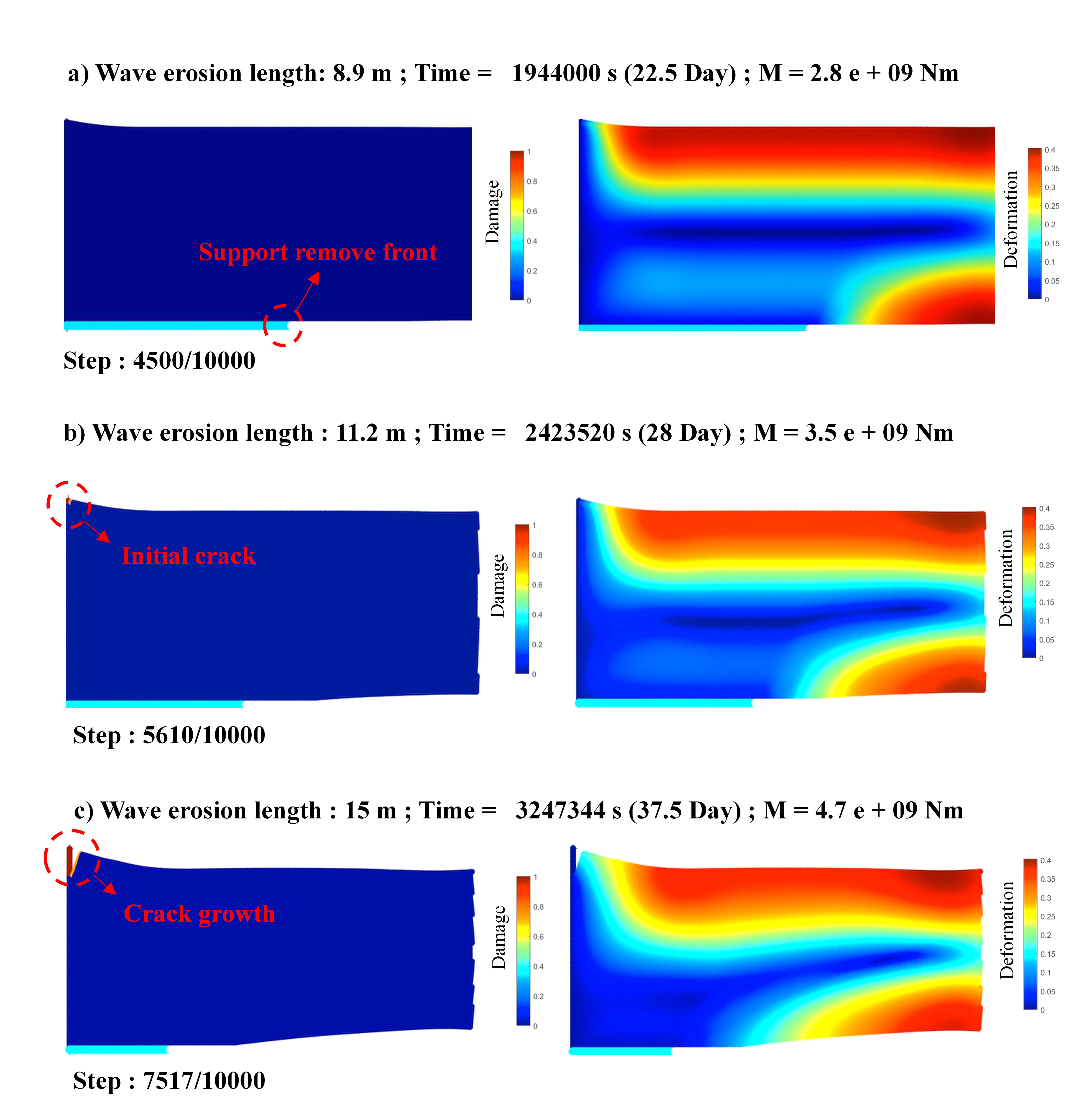}
  \end{center}		
  \caption{The progressive failure of ice shelf front under wave erosion}
  \label{fig:23}
\end{figure*}

This comparison serves two crucial purposes. First, it validates the accuracy of the PD model
in a well-understood, linear-elastic regime. Second, it establishes that the model possesses
the necessary sensitivity to reliably investigate scenarios beyond the scope of analytical theory,
 particularly where fracture, damage, and large deformations become dominant. Thus, the agreement
 shown here provides a solid foundation for using the PD method to study the subsequent stages
 of crack initiation and propagation in the foot-loosening process.

Fig. \ref{fig:23} chronologically illustrates the progressive failure of an ice shelf front under wave erosion, simulated using the peridynamic method. The sequence captures the transition from a state of stress concentration to macroscopic fracture. Fig. \ref{fig:23}(a)) illustrates the stress Concentration Following Support Loss. After the removal of basal support over an initial erosion length of 8.9 m (simulating 22.5 days of wave action), the ice shelf behaves as a cantilever. A significant bending moment ($\text{M}=2.8\times {{10}^{9}}~\mathrm{N\cdot m}$) develops, generating a zone of high tensile stress on its upper surface, as visually indicated. This stage establishes the mechanical precondition for fracture. Fig. \ref{fig:23}(b)) illustrates the crack Initiation. With continued erosion extending to 11.2 m (28 days), the bending moment increases to $3.5\times {{10}^{9}}~\mathrm{N\cdot m}$ . The accumulated tensile stress exceeds the ice's strength, leading to the nucleation of a discrete crack at the location of maximum stress concentration. This marks the transition from diffuse damage to a localized fracture. Fig.~\ref{fig:23}(c)) illustrates the crack Propagation. Further erosion to 15 m (37.5 days) and a corresponding rise in bending moment to $4.7\times {{10}^{9}}~\mathrm{N\cdot m}$ drive the stable growth of the crack. The fracture propagates inland, following the path of high tensile stress. This stage demonstrates how sustained wave forcing and the resulting increase in applied moment directly control the extent of fracture advancement prior to final calving.

\begin{figure}
  \begin{center}
       \includegraphics[height=4.0in]{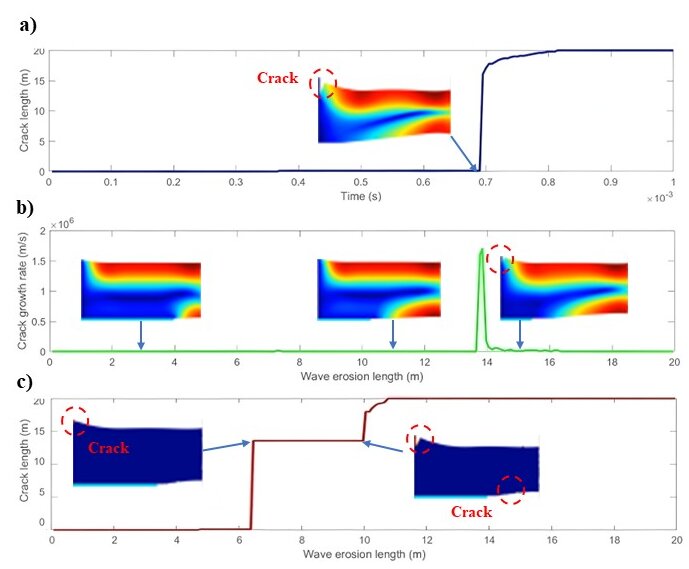}
  \end{center}   				
\caption{(a) Ice shelf fracture evolution induced by wave erosion,
  (b) The relationship between crack growth rate and wave erosion length, and (c) The relationship
  between crack length and wave erosion length}
  \label{fig:24}
\end{figure}

Fig. \ref{fig:24}(a) presents the simulated fracture evolution induced by wave erosion, illustrating the characteristic progression of the foot-loosening mechanism. Crack initiation is not immediate but occurs after a period of sustained stress accumulation. Following initiation, the crack propagates through the ice thickness primarily under tensile opening, with its length increasing progressively over time. The propagation rate is not constant but varies in response to the applied cyclic bending moment, advancing preferentially during phases of peak tensile stress within each wave cycle. Unlike gravity-driven collapse, fracture in this scenario is controlled by repeated stress cycling rather than monotonic loading, leading to gradual damage accumulation before crack onset.

Fig. \ref{fig:24}(b) illustrates the relationship between crack growth rate and wave erosion length derived from the simulations.
The crack growth rate remains nearly constant until a critical erosion length is reached, beyond
which it exhibits a marked decline. This trend indicates a shift in the dominant fracture mechanism, transiting from a stable,
propagation-controlled regime to one where the loss of basal support and altered stress distribution
begin to inhibit further crack advancement.

Fig. \ref{fig:24}(c) illustrates the relationship between crack length and wave erosion length,
a critical linkage in the wave-driven calving process. The data indicate that crack length
exhibits a positive, non-linear correlation with increasing erosion length.
Initially, crack growth remains limited as the developing notch primarily alters the local geometry.
Beyond a threshold erosion length, however, the crack undergoes accelerated propagation.
This transition occurs because the enlarged basal notch significantly amplifies tensile
bending stresses at the ice shelf front, creating a mechanically unstable configuration
that favors rapid fracture extension. This relationship underscores the role of progressive
geometric weakening as a primary driver of instability. The findings demonstrate that the final
calving size is not solely determined by instantaneous stress but is intrinsically coupled
to the cumulative history of frontal erosion. Consequently, monitoring and predicting
wave-induced notch development becomes essential for assessing the imminence and potential
magnitude of calving events associated with the foot-loosening mechanism.

\begin{figure}[ht]
  \begin{center}            				
      \includegraphics[height=4.0in]{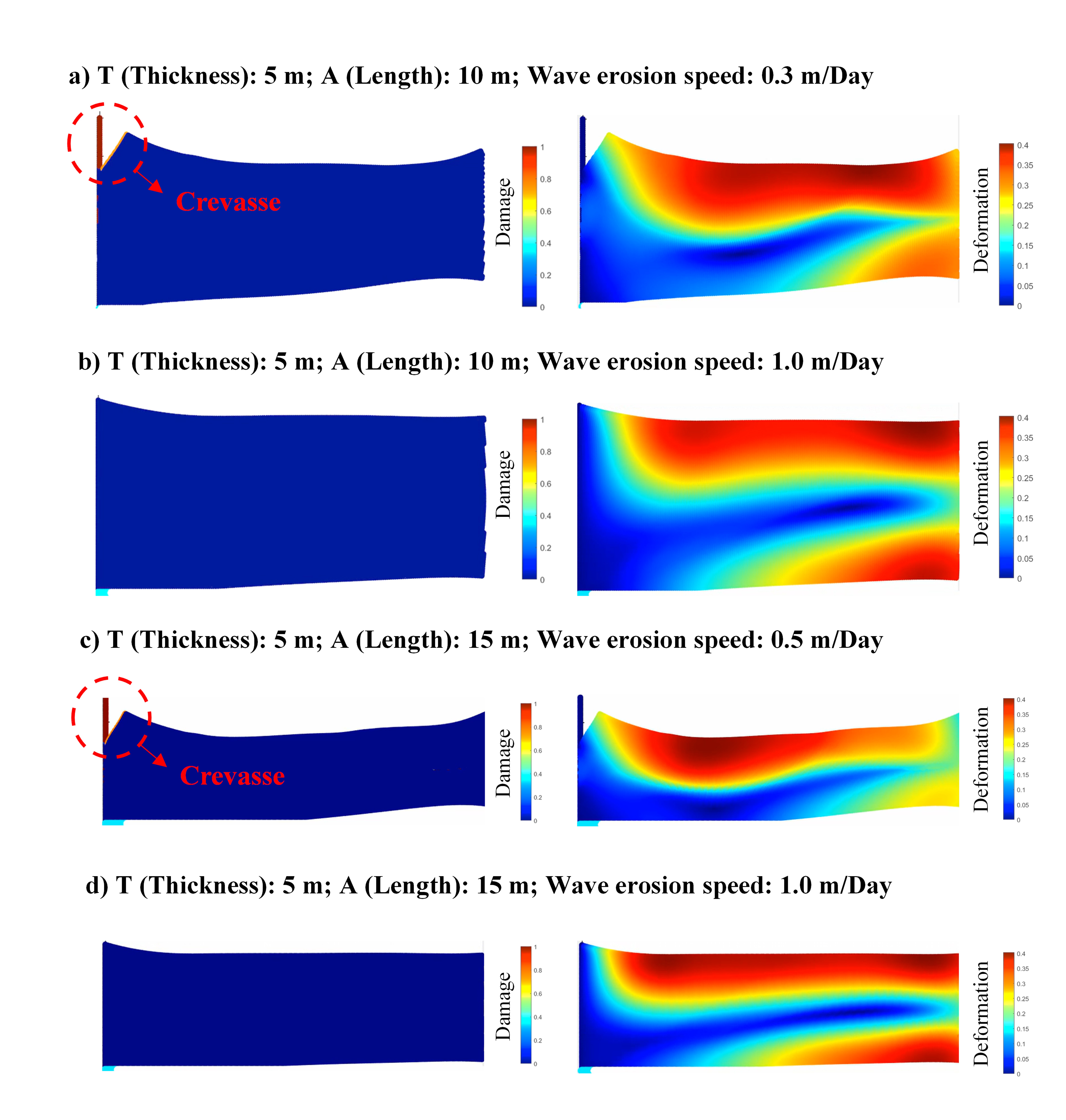}
  \end{center}			
  \caption{Crack length evolution characteristics with varying wave erosion speed}
  \label{fig:27}
\end{figure}

The influence of wave erosion rate on fracture evolution is summarized in Fig.~\ref{fig:27}. At lower erosion rates, stress has sufficient time to redistribute, allowing damage to accumulate over a broader region before crack initiation. This results in more distributed cracking and less localized fracture paths. In contrast, higher erosion rates lead to more localized stress concentration near the ice front, triggering earlier crack initiation and faster crack propagation along a narrow zone. These results indicate that wave erosion rate plays a critical role in controlling both the timing and morphology of foot-loosening-induced fractures.

The simulated behavior aligns closely with the conceptual foot-loosening model, confirming that repeated wave-induced bending can generate sufficient tensile stress to drive crack propagation over scales relevant to ice shelf destabilization.

Overall, Scenario 2 demonstrates that wave-induced bending alone can generate sufficient tensile stress to trigger fracture and drive large-scale crack propagation in floating ice shelves. The results provide a mechanistic link between continuous wave action and discrete calving events, supporting the hypothesis that foot-loosening is an effective pathway for ice shelf destabilization.

\section{Conclusions}
\label{sec:6}

In this study, we employ a peridynamics (PD) framework to resolve the physical processes governing ice-shelf calving induced by wave erosion, basal melting, and buoyancy-driven flexure, with a focus on two sequential failure mechanisms frontal collapse and foot loosening. \sout{Beyond advancing} \textcolor{black}{In addition to} fracture mechanics, this work directly addresses a critical limitation in current ice-sheet and sea-level-rise models: the lack of physics-based calving laws capable of representing wave-driven fracture and buoyancy-controlled failure at ice-shelf fronts.

The key findings and broader impacts are summarized as follows:

\begin{enumerate}
\item
Physics-based calving processes relevant to large-scale ice-sheet models.
The PD model captures crack initiation, microcrack coalescence under gravity-dominated and wave-induced cyclic loading, crevasse propagation, and ice-block fragmentation within a unified nonlocal framework. The crack-tip stress fields, validated against finite-element solutions and analytical beam-theory predictions, provide a mechanistic description of stress redistribution and fracture evolution that can inform physically grounded calving parameterizations, replacing empirical or threshold-based formulations commonly used in ice-sheet models.

\item
Wave erosion as a missing driver of ice-shelf destabilization and mass loss.
We demonstrate that wave erosion exerts a dual and previously underrepresented control on calving behavior. Progressive basal notch formation promotes localized frontal collapse under self-weight, while cyclic wave-induced bending generates large-scale tensile stresses that drive buoyancy-controlled flexural failure (foot loosening). This mechanism produces fractures extending hundreds of meters along the ice-shelf front, effectively preconditioning large ice blocks for detachment. Such processes are not explicitly represented in current sea-level-rise projections, suggesting that wave forcing may contribute to underestimated calving rates in vulnerable ice shelves.

\item
Threshold behavior and scaling laws for calving parameterization.
Crack and crevasse propagation exhibit strong nonlinear dependence on erosion rate, erosion length, and bending flexural moment. Once a critical erosion length is exceeded, fracture accelerates rapidly due to amplified bending stresses, while higher erosion rates favor earlier crack initiation and more localized failure. These threshold behaviors provide a quantitative basis for developing scalable calving criteria that link ocean-wave energy and shelf geometry to calving frequency and magnitude in continental-scale ice-sheet models.
\end{enumerate}

Overall, this study establishes a direct pathway for incorporating mechanistically informed, wave-driven calving laws into next-generation ice-sheet and sea-level-rise models. By explicitly resolving how ocean forcing translates into fracture initiation and large-scale ice-shelf failure, the proposed PD framework improves confidence in projections of ice-shelf stability, grounded ice discharge, and future sea-level rise under a warming and increasingly energetic ocean.

\bigskip
\section*{Statements of conflict interests}
The authors declare that they have no conflict of interest.

\bigskip
\section*{Acknowledgements}
S.L. and W. L. thank the Peder Sather Grant from the Peder Sather Center for Advanced Study at
the University of California, Berkeley
for supporting this research.
Y. S. performed the research at the Harbin Engineering University.
These supports are greatly appreciated.

\newpage



\bibliographystyle{spmpsci}      
\bibliography{ref}

\newpage
\section*{Appendix}

In this Appendix, we present the finite element modeling of a two-dimensional ice shelf
stress distribution as well as the displacement distribution in Fig. 22 and 23.

To supplement the quantitative analysis, a complete set of contour plots is provided to visualize the mechanical response 
across $R/t$ ratios ranging from 0.1 to 0.9. Figure 22 shows the redistribution of longitudinal tensile stress ($S_{11}$), 
highlighting the effectiveness of an increased fillet radius in mitigating local stress peaks. Figure 23 
presents the corresponding displacement fields, confirming the stability of the deformation pattern under gravitational loading. 
These results show that as the base profile becomes smoother due to erosion, the tensile stress is reduced, 
which lowers the risk of ice shelf collapse.

\begin{figure*}[ht]
  \begin{center}            				
      \includegraphics[height=1.7in]{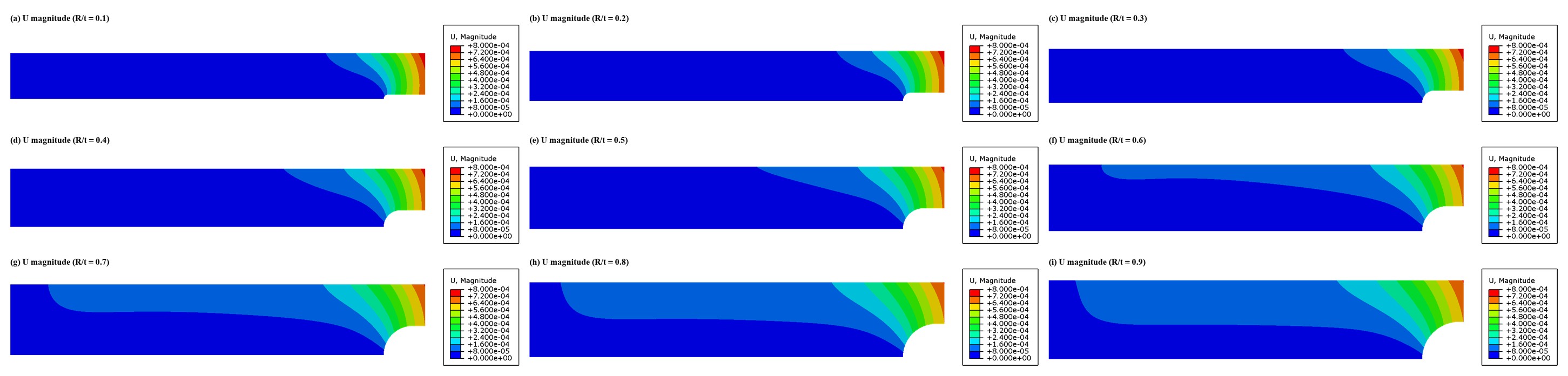}
  \end{center}		
  \caption{ 
  Finite element modeling results of stress distribution
  over a two-dimensional
  ice-shelf model.
  Finite element modeling results of displacement distribution over a two-dimensional
  ice-shelf model.
  Complete set of $S_11$ tensile stress distributions. The matrix displays the evolution 
  of longitudinal tensile stress across the ice shelf overhang for R/t ratios ranging from 0.1 to 0.9, 
  illustrating the mitigation of stress at the transition zone.
}	
  \label{fig:appendix1}
\end{figure*}

\begin{figure*}[ht]
  \begin{center}            				
      \includegraphics[height=1.7in]{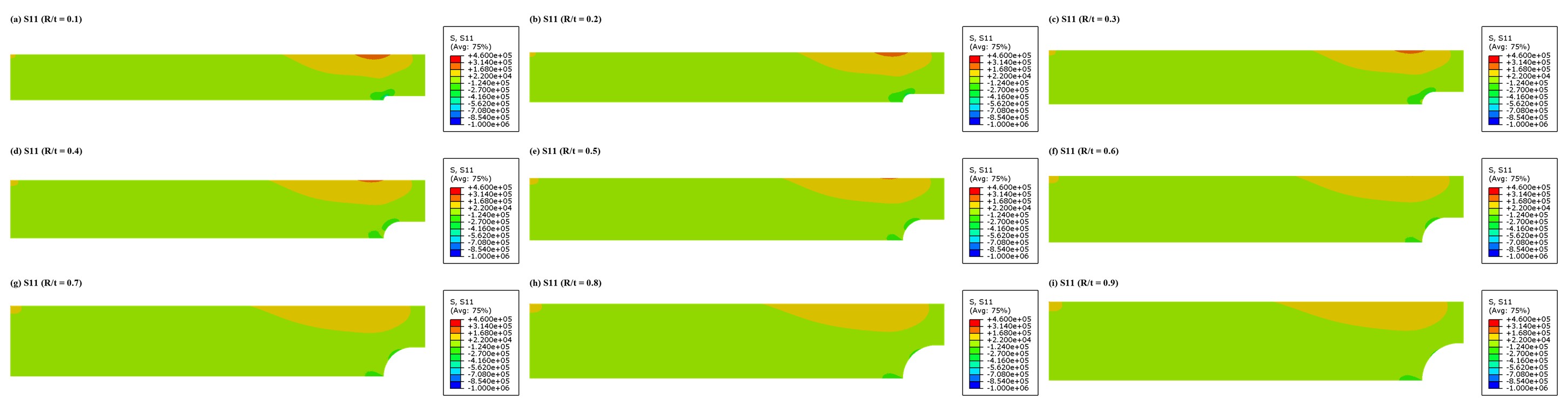}
  \end{center}		
  \caption{ Finite element modeling results of displacement profile
  over a two-dimensional
  ice-shelf model.
  Complete set of total displacement magnitude distributions. 
  The matrix presents the global deformation patterns for R/t ratios ranging from 0.1 to 0.9 under gravitational loading.
  }		
   \label{fig:appendix2}
\end{figure*}

\end{document}